\documentclass[9pt]{IEEEtran}

\newtheorem{remark}{Remark}

\usepackage{cite}
\usepackage{amsmath,amssymb}
\usepackage{graphicx,subfig}
\usepackage{hyperref}

\hypersetup{
  colorlinks   = true, 
  urlcolor     = blue, 
  linkcolor    = blue, 
  citecolor   = red 
}

\usepackage{docmute}
\usepackage[table]{xcolor}

\usepackage{tabularx}
\newcolumntype{L}[1]{>{\raggedright\let\newline\\\arraybackslash\hspace{0pt}}m{#1}}
\newcolumntype{C}[1]{>{\centering\let\newline\\\arraybackslash\hspace{0pt}}m{#1}}
\newcolumntype{R}[1]{>{\raggedleft\let\newline\\\arraybackslash\hspace{0pt}}m{#1}}

\usepackage{MathDefs}

\usepackage{algorithm}
\usepackage{algpseudocode} 

\usepackage{verbatim}
\usepackage{tikz}

\makeatother

\begin{document}
\title{A Novel Slip-Kalman Filter to Track the Progression of Reading Through Eye-Gaze Measurements}

\author{
S. Bottos and 
B. Balasingam, {\em Senior Member, IEEE}
\thanks{
The authors are with the 
Department of Electrical and Computer Engineering, 
University of Windsor, 401 Sunset Avenue,
Windsor, Ontarion, N9G 3P4, Canada.
E-mail:  \{bottos,singam\}@uwindsor.ca,
Contact TP: +1(519) 253-3000 ext. 5431, 
Fax: +1(519) 971-3695
} 
}

%

\maketitle

\begin{abstract}
In this paper, we propose an approach to track the progression of eye-gaze while reading a block of text  on computer screen. 
The proposed approach will help to accurately quantify reading,
e.g., identifying the lines of text that were read/skipped and estimating the time spent on each line,
 based on commercially available inexpensive eye-tracking devices.
 The proposed approach is based on a novel {\em slip Kalman filter} that is custom designed to track the progression of reading. 
 The performance of the proposed method is demonstrated using 25 pages eye-tracking data collected using a commercial desk-mounted eye-tracking device. 
\end{abstract}


\section{Introduction}
Eye-gaze tracking, in particular its applications in reading-pattern analysis, has been a topic of interest for over a century. Pioneers such as Louis Javal \cite{roper2007louis} and Edmund Huey \cite{huey1908psychology} paved the field's foundation by searching for connections between an individual's reading patterns and their emotions, fatigue level, or mental condition. Indeed, the analysis of eye-gaze fixation patterns has since been used to detect disorders such as depression, anxiety \cite{sanchez2013attentional, armstrong2012eye}, and autism \cite{neumann2006looking, klin2002visual}. The study of reading itself however has continued to prove quite difficult in practice due to the complex nature of such a simple action \cite{rayner1997understanding, drieghe2005eye}. 
Recent works have addressed the issue of translating an individual's eye-gaze fixation points to an accurate interpretation of their reading patterns and habits with modern solutions\cite{paeglis2006maximizing, hochreiter1997long, mozaffari2018reading}, yielding results far more quickly and with unprecidented accuracy than ever before. There is one caveat however, while these modern solutions yield excellent results, they are still reliant on expensive hardware and regulated environments which restrict lighting, head movement, and distance from the text being read much like their primitive counterparts.

\par
The objective of this paper is to contribute to the research and development of methods which are not only capable of extracting useful, accurate information from messy, noise-corrupt eye-gaze fixation points collected during reading, but are also robust enough to do so in a minimally restrictive, more natural environment and without the need for expensive hardware. Specifically we seek to develop an approach to draw conclusions, which include but are not limited to, lines which have read or not-read, lines which were read multiple times, and how much time was spent on reading each line. The intent is to make accurate reading-pattern tracking more accessible using inexpensive, non-invasive hardware without requiring the subject limit their movements in such a way that hinders natural behaviour. In this regard, the work presented in this paper builds on two recent works by the authors \cite{bottos2019approach, bottosFusion}. In \cite{bottos2019approach}, an approach developed by the authors, creatively given the name ``Line Detection System (LDS)'' was developed to detect the line being read at any given instant during reading; in \cite{bottosFusion}, a least squares estimator is developed to compliment the LDS in tracking the progression of reading along each line. 

\par
The LDS approach in \cite{bottos2019approach} relies on discrete hidden Markov models (HMM) to identify lines.  
Discretization is accomplished by assigning data whose $y$-coordinates were filtered to reduce noise an ``observed line number'' based on where each eye-gaze fixation point was measured relative to the grid boundaries. 
As expected, it was found that the HMM based LDS takes {\em transition time} when the reader's eye-gaze progresses from one line to the next. This coupled with the fact that the system is prone to mis-classifying large sequences of eye-gaze fixation points as incorrect observed line numbers results in {\em line detection errors}. While the LDS in its current state is able to detect, with typically above 90\% accuracy, a reader's line-by-line progression, its accuracy is slightly volatile due to the presence of line detection errors in some instances. The system is also incapable of real-time smoothing of eye-gaze fixation points, nor is it functional in a situation where the number of lines is not prior knowledge. Finally, adequate noise-reduction, especially with that associated with the collected $x$-coodinates, presents an interesting challenge - discussed in \cite{bottosFusion} and, to a lesser degree, \cite{bottos2019approach} - which has not yet been attended to sufficiently in previous works. 

In this paper, we propose an alternative designed to address the flaws identified with previous approaches. Rather than using a HMM as a classifier to estimate line progression - as with the LDS - we propose the use of a modified Kalman filter to detect when the reader's eye-gaze continues from one line to another. Particularly, we report the filtered {\em horizontal velocity} of the Kalman filter as a superior indicator of new lines. The proposed Kalman filter is termed {\em slip Kalman filter (Slip-KF)} based on its ability to reset when progression from the end of one line to the beginning of another is detected. Furthermore, the Slip-KF offers vastly improved estimation of $x$-coodinates when compared to the previos least squares method. 
\par
The remainder of this paper is organized as follows: in Section \ref{sec:probdef} a review is given on the standard Kalman Filter, and its inability to perform properly when applied to reading-pattern detection is highlithed. In Section \ref{sec:PA}, we present an improved version of the standard Kalman filter which is capable of maintaining its accuracy in an environment where erratic movement is common, such as during reading. 
The paper is concluded in Section \ref{sec:conclusions}. 

\section{Problem Definition}
\label{sec:probdef}
The authors intend to utilize the power of the Kalman filter to smooth and interpret noisy eye-gaze fixation point data collected while reading. Eye-gaze data collected during reading tends to be difficult to interpret, due to the presence of noise intro- duced as the result of innacuracies in detection hardware, intrinsic characteristics of eye-gaze which dictates that eye movement itself is inherently erratic, and head movement of the individual being tracked. During reading, specifically, the eye does not scan text being read at an equal rate. Progression from the beginning of one line to its end occurs as a sequence of quick and uneven saccades, while the progression from one line to a new line altogether manifests as a quick saccade coupled with a drastic change in position - resulting in a rapid change in velocity and acceleration. We will begin with a brief review of the standard Kalman filter prior to discussing the proposed modifications.

Let us model the progression of eye-gaze during a reading activity as the following $4 \times 1$ {\em state vector} 
\begin{eqnarray}
\bx(k) &=& [ x(k) \,\,\,\, \dot x(k) \,\,\,\, y(k) \,\,\,\,  \dot y(k)  ]^{\rm T} 
\end{eqnarray}
where 
$x(k)$ is the gaze-displacement in the x-direction, 
$y(k)$ is the gaze-displacement in the y-direction, 
$\dot x(k)$ is the {\rm rate of gaze-displacement} (velocity) in the x-direction, and
$\dot y(k)$ is the rate of gaze-displacement (velocity) in the y-direction. 

The state vector above is modeled to undergo the following {\em process model}
\begin{eqnarray}
\bx(k+1) = \bF \bx(k) + \bGamma \bv(k) 
\label{eq:process}
\end{eqnarray}
where the elements of the $4\times1$ vector $\bv(k) $ are assumed to be {\em zero-mean Gaussian noise with unity standard deviation}, 
\begin{eqnarray}
\bF= 
\begin{bmatrix}
1 & \Delta T & 0 & 0  \\
0 & 1 & 0 & 0  \\
0 & 0 & 1 & \Delta T  \\
0 & 0 & 0 & 1 \\
\end{bmatrix}, \quad
\bGamma = 
\begin{bmatrix}
\frac{1}{2} \Delta T^2 & 0 \\
\Delta T & 0\\
0 & \frac{1}{2} \Delta T^2 \\
0 & \Delta T
\end{bmatrix}
\begin{bmatrix}
q_x & 0\\
0 & q_y 
\end{bmatrix}
\end{eqnarray}
where $\Delta T$ is the sampling time that is assumed to be a constant. 
The values of $q_x$ and $q_y$ need to be selected based on realistic insight about the application. 
For example, in the case of reading, we will have $q_x > q_y. $
The {\em process noise covariance}, $\bQ$, can be shown to be \cite{bar2004estimation}
\begin{eqnarray}
\bQ &=& E\left\{\bGamma \bv(k) \bv(k)^{\rm T} \bGamma^{\rm T}  \right\} \\
&=&
\begin{bmatrix}
\frac{1}{4}\Delta T^4 q_x & \frac{1}{2}\Delta T^3 q_x& 0 & 0 \\
\frac{1}{2}\Delta T^3 q_x& \Delta T^2 q_x& 0 & 0 \\
0 & 0 & \frac{1}{4}\Delta T^4 q_y & \frac{1}{2}\Delta T^3q_y \\
0 & 0 & \frac{1}{2}\Delta T^3 q_y& \Delta T^2 q_y \\
\end{bmatrix}
\end{eqnarray}
where $\bQ$ is the covariance of the process noise.

\begin{remark}
An approach to select $q_x$ and  $q_y$ is as follows \cite{bar2004estimation}.
Let us assume that a text of 25 equally spaced lines on a screen measuring $8.5$ inches sideways and $11$ inches from top to bottom. 
Assuming that a person requires, on average, 10 seconds to finish reading one line, the average horizontal velocity of eye-gaze on a per-line basis, taking left-to-right progression as positive, is
\begin{eqnarray}
\dot x(k)= \frac{1}{10} \times 8.5 = 0.85  \,\, {\rm inches/sec}
\end{eqnarray}
Now, allowing 1\% change (noise) in velocity, a reasonable value to assign to $q_x$ is obtained as follows
\begin{eqnarray}
\sqrt{\Delta T^2 q_x} &=& 0.85 \times \frac{1}{100} = 0.0085 \nonumber \\
q_x & =& \left( 0.0085/\Delta T \right)^2 
\end{eqnarray}
Considering that it takes 10 seconds to move from one line to another, the vetical velocity on a per-page basis, taking downward progression as positive, is
\begin{eqnarray}
\dot y(k)= \frac{1}{10} \times \frac{11}{25} = 0.044  \,\, {\rm inches/sec}
\end{eqnarray}
Assuming a similar 
\begin{eqnarray}
\sqrt{\Delta T^2 q_x} &=& 0.0044 \times \frac{1}{100} = 0.000044 \nonumber \\
q_x & =& \left( 0.000044 /\Delta T \right)^2 
\end{eqnarray}
Note that the horizontal progression was measured per-line while the vertical progression was measured per-page. Note also that the unit of measurement and sampling time may differ depending on the device used to obtain measurements. The previous example was meant to illustrate a general technique rather than act as a rigid approach. 
\end{remark}

The eye-gaze (or more accurately eye-gaze fixation point) measurements are the noisy observations of $x(k)$ and $y(k)$, denoted in vector form as
\begin{eqnarray}
\bz(k) &=& [ z_x(k) \,\,\,\, z_y(k) ]^{\rm T} 
\end{eqnarray} 
Let us define the {\em measurement model}, $\bz(k)$ that match the eye-fixation observations to the state vector $\bx(k)$ or reading as follows
\begin{eqnarray}
\bz(k) = \bH \bx(k) + \bw(k) 
\label{eq:meas}
\end{eqnarray}
where
\begin{eqnarray}
\bH= 
\begin{bmatrix}
1 & 0 & 0 & 0  \\
0 & 0 & 1 & 0 \\
\end{bmatrix}, \quad
\end{eqnarray}
and the elements of the $2\times1$ vector $\bw(k) $ are assumed to be {\em zero-mean Gaussian noise with standard deviation} $\sigma_x$ for the $x$-coordinate measurement and $\sigma_y$ for the $y$-coordinate measurement.
For simplicity, we assume the measurement noise to be uncoorelated in x, y directions. 
Consequently, the measurement model covariance matrix is written as 
\begin{eqnarray}
\bR = \begin{bmatrix}
\sigma_x^2& 0  \\
0  & \sigma_y^2
\end{bmatrix}
\end{eqnarray}

Given an initial estimate for $\bx(0|0)$ and $\bP(0|0)$ the Kalman filter \cite{bar2004estimation} can be used to recursively obtain the updated estimate of $\hat \bx(k+1|k+1)$ and the estimation error covariance $\bP(k+1|k+1)$  as each new measurement $\bz(k)$ arrives, for $k = 0:K-1.$
Algorithm \ref{alg:regularKF} summarizes one recursive step of the regular KF. 

\begin{algorithm}
\caption{Regular Kalman Filter, Single Iteration}
\label{alg:regularKF}
\begin{algorithmic}[1]
\State {\bf input:}  $\bx(k|k)$, $\bP(k|k)$, $\bz(k+1)$
\State State prediction: $\hat \bx(k+1|k)=\bF\bx(k|k) $
\State State prediction Cov: $ \bP(k+1|k)=\bF\bP(k|k)\bF^{\rm T}+\bQ $ 
\State Innovation Cov: $ \bS(k+1)=\bR+\bH\bP(k+1|k)\bH^{\rm T}$
\State Measurement prediction: $\hat z(k+1|k)=  \bH \hat \bx(k+1|k)$ 
\State Measurement Residual: $ \nu (k+1)=  z(k+1)-\hat z(k+1|k) $ 
\State State Est: $ \hat \bx(k+1|k+1) = \hat \bx(k+1|k)+ \bW(k+1)v(k+1)  $
\State State Est. Cov: $\bP(k+1|k+1) =  \bP(k+1|k)- \bW(k+1)\bS(k+1)\bW(k+1)^{\rm T}  $ 
\State \bf{return} $\hat\bx(k+1|k+1)$, $\bP(k+1|k+1)$
\end{algorithmic}
\end{algorithm}

\begin{remark}[Filter initialization]
Algorithm \ref{alg:regularKF} needs initial values; i.e, $\bx(0|0)$ and $ \bP(0|0)$ needs to be computed. 
A simple way to obtain the initial estimate is the two-point initialization method \cite{bar2004estimation}. 
Given the first two measurements 
$z(1) = [z_x(1), z_y(1)]T$ and $ z(2) = [z_x(2), z_y(2)]T$, the initial estimate is obtained as 
\begin{eqnarray}
\bx(0|0) = \left[z_x(2), \frac{z_x(2)-z_x(1)}{\Delta T}, z_y(2), \frac{z_y(2)-z_y(1)}{\Delta T} \right]
\end{eqnarray}
and the filter covariance is initialized as 
\begin{eqnarray}
\bP(0|0) = \gamma \bI_4
\end{eqnarray}
where $\gamma$ is an appropriately large coefficient and $ \bI_4$ is the $4 \times 4$ identity matrix. 
\end{remark}

\begin{figure}[htbp]
\begin{center}
\includegraphics[width=.45\textwidth]{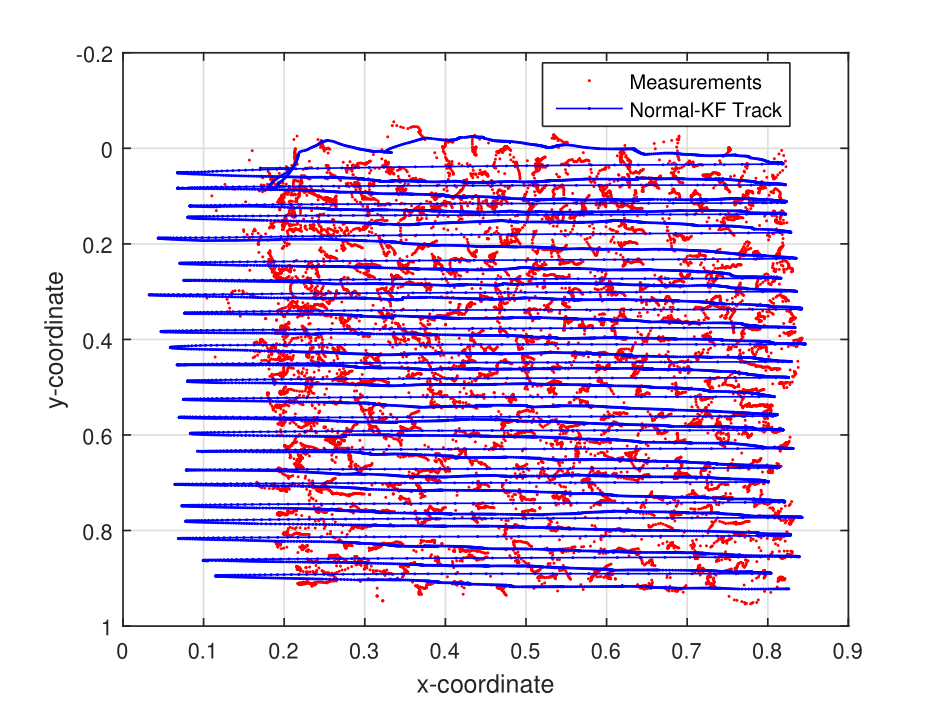}
\caption{{\bf Tracking eye-gaze while reading using normal Kalman filter.} The standard Kalman filter is unsuitable for use with eye-gaze data collected during a reading activity, due to the presence of very pronounced over-estimation at each line-change.}
\label{fig:regularKF}
\end{center}
\end{figure}

Figure \ref{fig:regularKF} illustrates the performance of a regular Kalman filter, when used to eliminate noise from eye-gaze data collected during a reading activity, from one sample page of data. The measurements $\bz(k)$ are shown as red `*" and the estimated coordinates are shown as a blue line. 
These measurements were obtained by tracking an individual as they read 25 pages worth of text, each page having 25 lines per page. More details on the experimental setup used to take this measurement can be found in Section \ref{sec:results} as well as in \cite{bottos2019approach}. 

The standard Kalman filter is not designed to track motion which behaves as erratically as one's eye-gaze during reading. It can be noted that at the end of each line, the drastic change in state results in over-estimation of the true position. After a short amount of time, somewhere around mid-line, the filter re-establishes a steady state and resumes accurate estimation, only to be disturbed anew at the line's completion. This behaviour, however, also presents an opportunity to develop a method of self-correction, which is the basis of the Slip-Kalman Filter.

\section{Proposed Approach}
\label{sec:PA}
The title for the proposed modification to the Kalman filter is inspired by the slip gear \cite{hughes1910slip} used commonly in modern robotic applications. When a load exceeds a certain threshold, the slip gear is designed to {\em slip back}, effectively avoiding failure. Similarly, the proposed {\em Slip-KF} is designed to {\em re-initialize} when a certain threshold is exceeded. 
Particularly, the Slip-KF is designed to reset whenever a {\em line change} is detected based on the eye-gaze measurements - due to the fact that the state-space model defined by \eqref{eq:process} and \eqref{eq:meas} does not account for the near-instantaneous change in position, velocity, and acceleration triggered by the flick of one's eyes from one side of the page to the other.  The question is, how can a line change be detected such that the filter can be reset accordingly? 

\def\LTR{This is only true for language scripts that are written left-to-right, such as English.
However, the observation is true -- with the direction of velocity reversed -- for language scripts that are written right-to-left, such as Arabic. 
}

\begin{figure}[h]
\begin{center}
\subfloat[][NIS of the regular KF]{{\includegraphics[width=.45\textwidth]{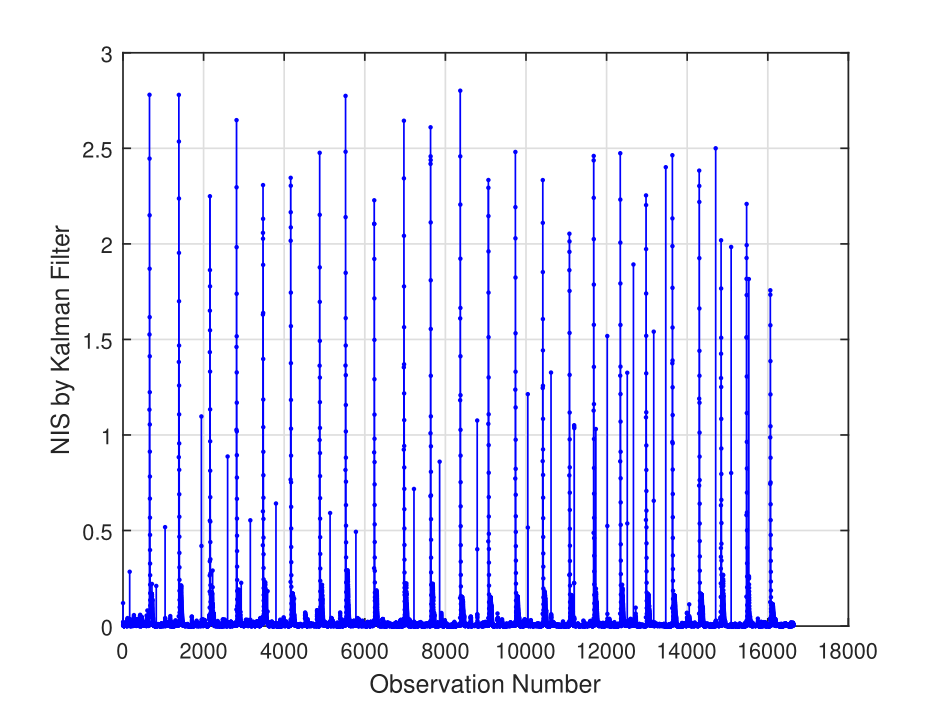}}
\label{nis}}
\\
\subfloat[][Velocity $\dot x(k)$, in text-width/sec, of the regular KF]{{\includegraphics[width=.45\textwidth]{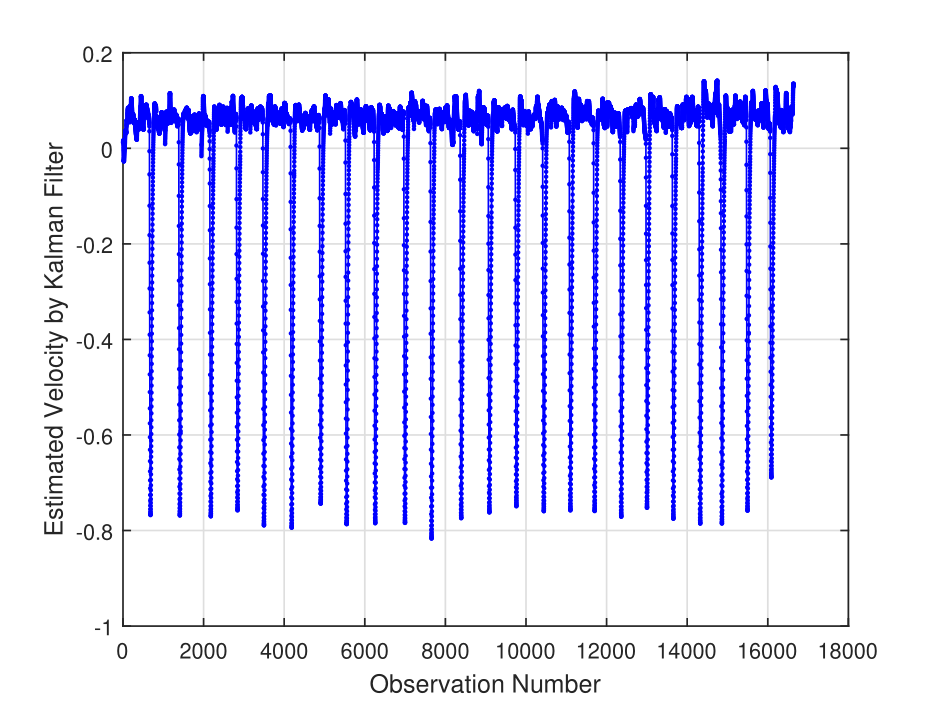}}
\label{vel}}
\caption{
{\bf Possible indicators of a new line.}]
Both NIS and the estimated velocity were candidates for new line indicators.
However, the velocity indicator simply outperforms NIS as indicator of a new line while reading. 
}
\label{fig:NISvsVel}
\end{center}
\end{figure}

The Slip-KF is ``tuned'' to follow typical reading progression without the expectation of a line change. Whenever there is a line change, the sudden change in state is analogous to the filter being ``overloaded'' in the slip-gear analogy. Such an overload can be observed in two filter parameters. 
\begin{itemize}
\item {\em Normalized Innovation Squared (NIS).}
It can be noticed that the NIS  \cite{bar2004estimation} spikes each time the filter is {\em stressed} -- which is likely to happen whenever the eye-gaze moves faster than that is expected by the filter, or whenever there is a vast discrepancy between the estimated measurement and the true measurement. Figure \ref{fig:NISvsVel}\subref{nis} illustrates the values of NIS associated with a regular KF during reading, from one sample page of data. It was observed that NIS spikes occur quite frequently due to the noisy nature of the eye-gaze
data, and thus this metric proved unreliable in terms of line change detection.
\item {\em Velocity along the line $\dot x(k)$}.  
Reading is characterized, majoritively, by left-to-right eye travel\footnote{\LTR} as the reader scans each line of text. It is reasonable to treat the {\em velocity} of such left-to-right eye-gaze-movement as fairly constant, even though in reality reading occurs in quick saccades. However, when a reader reaches the end of one line and adjusts their gaze from right to left to begin the next, a spike in velocity {\em in the opposite direction} compared to the typical left-to-right eye travel is observed. Figure \ref{fig:NISvsVel}\subref{vel} illustrates this phenomenon, in which each spike in velocity is clearly defined. Such a pronounced, consistent difference allows us to establish a threshold for new line detection. 
\end{itemize}

\begin{remark}[Useful observations from the KF]
Some additional interesting observations can be made form Figure  \ref{fig:NISvsVel}\subref{vel}:
it can be noted that for this particular sample page, the average velocity is approximately 1/3 of 0.2 $\approx 0.067$; from this, it can be concluded that the person took $1/0.067 \approx 15$ seconds to finish one line of text. 
\end{remark}


A summary of the proposed Slip-KF is given in Algorithm \ref{alg:slipKF}. The Slip-KF monitors the estimated velocity at each instant, checking it against a pre-defined threshold. If this threshold is exceeded, rather than produce an estimate for this instant the Slip-KF re-initializes the state estimation covariance and state vector - essentially restarting the filter after each new line without sacrificing a drastic loss in momentum or convergence.
\begin{algorithm}
\caption{(Slip-Kalman Filter, Single Iteration)}
\label{alg:slipKF}
\begin{algorithmic}[1]
\State {\bf input:}  $\hat\bx(k|k)$, $\bP(k|k)$, $\bz(k+1)$
\State {\bf if}{ $\hat {\dot x}(k) < -0.5  $} {\bf then:}
\\ \quad\quad {\bf re-initialize:}
\\\quad\quad\quad {$\hat \bx(k+1|k+1)  = [  z_x(k+1), 0.2/3 ,  z_y(k+1), \hat {\dot y}(k|k) ]^{\rm T} $}
\\ \quad\quad\quad {$\bP(k+1|k+1) = \bP(0|0)$}
\\ \quad\quad\quad {\bf return} $\hat\bx(k+1|k+1)$, $\bP(k+1|k+1)$
\State \quad {\bf else:}
\\ \quad\quad Execute regular Kalman Filter (Algorithm \ref{alg:regularKF}) 

\end{algorithmic}
\end{algorithm}

\begin{remark}[Filter re-initialization]
The re-initialization step in Algorithm \ref{alg:slipKF} is performed as follows:
the $x,y$ states are re-initialized to measurements, i.e., $x(k+1|k+1) = z_x(k+1)$ and $y(k+1|k+1) = z_y(k+1)$;
the $x$-velocity is  re-initialized to typical reading rate of  $\dot x(k+1|k+1) = 0.2/3$ page-width/second and the 
$x$-velocity is re-initialized to its prior estimate  $\dot y(k+1|k+1) = \dot y(k|k)$ . 
\end{remark}

\section{Results}
\label{sec:results}

Eye gaze data were collected from a single test subject (a male in his twenties), using a Gazepoint GP3 \cite{gazept} desktop device. To begin logging data, the test subject was required to press the space key which would simultaneously cue the device to begin logging the $x$ and $y$ eye-gaze fixation coordinates, $\bz(k)$, at 64 Hz, and reveal a single line of text against a solid background, near the top of the display (in this case, a 1920$\times$1080 computer monitor) for which the device was calibrated. While the topmost line of text  - Line 1 - was displayed, each gaze point corresponding to this line were labeled with a ``1'' to allow for comparison between ground truths and predictions. Only a single line of text at a time was displayed, starting from line 1 at the top of the screen to line 25 at the bottom of the screen, during reading and tracking. Each new line of text was shown by pressing space when the end of each line of text had been reached, allowing ground truths to be recorded along with each eye-gaze fixation point. For more details about the collected data, please refer \cite{bottosFusion,bottos2019approach}.

\begin{figure}
\begin{center}
\includegraphics[width=.45\textwidth]{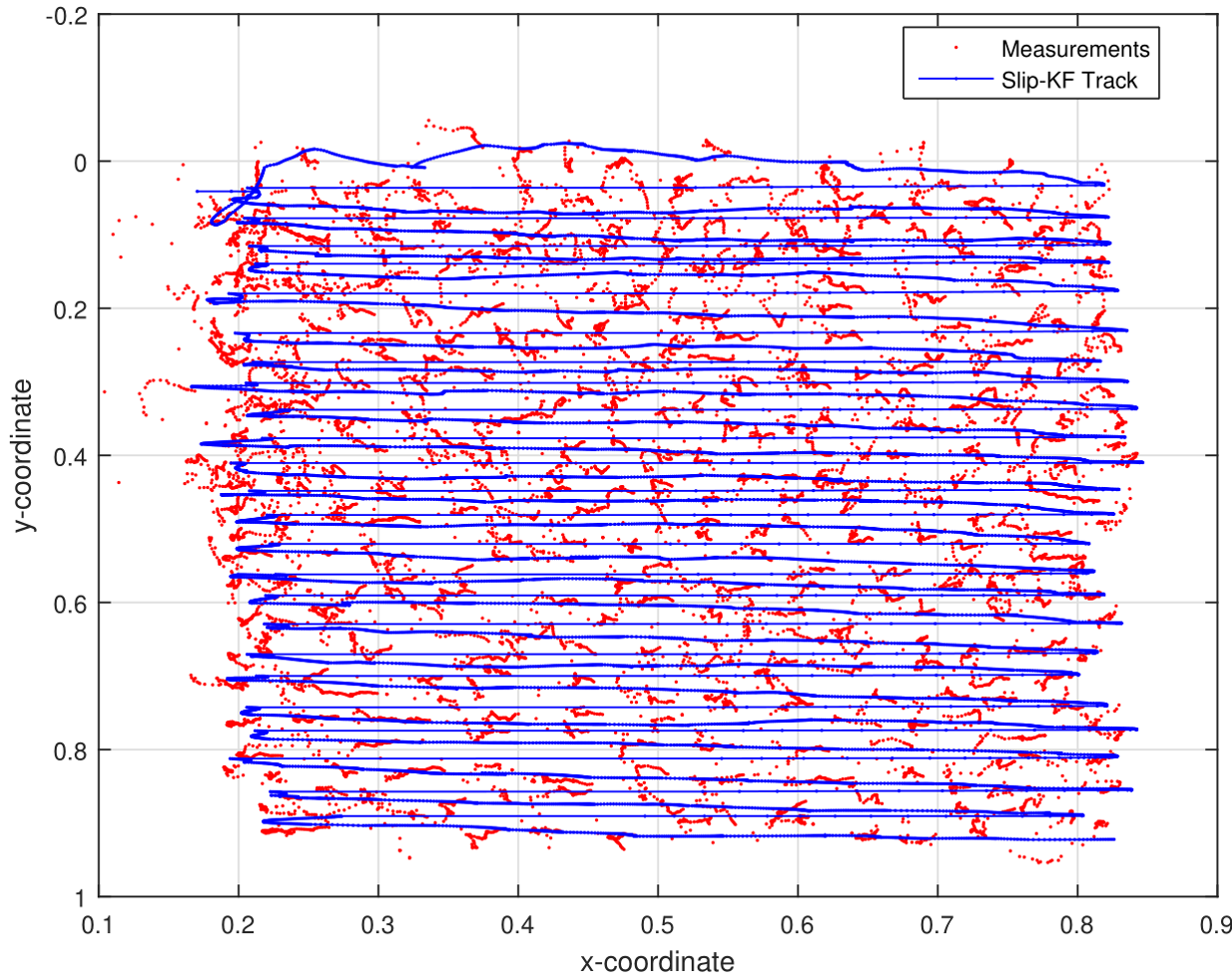}
\caption{
{\bf Eye-gaze tracking while reading with the proposed Slip-KF.}
Regular Kalman filter is able to follow the eye-fixation measurements along the line. However, when there is a line return (i.e., when the reader moves from the end of one line to the start of a new line) the filter is `thrown back' and takes time to recover. 
}
\label{fig:slipKF}
\end{center}
\end{figure}

Figure \ref{fig:slipKF} shows the result of the proposed Slip-KF while reading a sample page of the above mentioned experimental data. This figure must be viewed in comparison to  Figure \ref{fig:regularKF} which used the regular kalman Filter defined in Section \ref{sec:probdef};
The advantage of the Slip-KF is immediately clear - over-estimation is avoided, and effective noise reduction is achieve. 

\begin{figure}
\begin{center}
\subfloat[][True vs. predicted line]{\includegraphics[width=.45\textwidth]{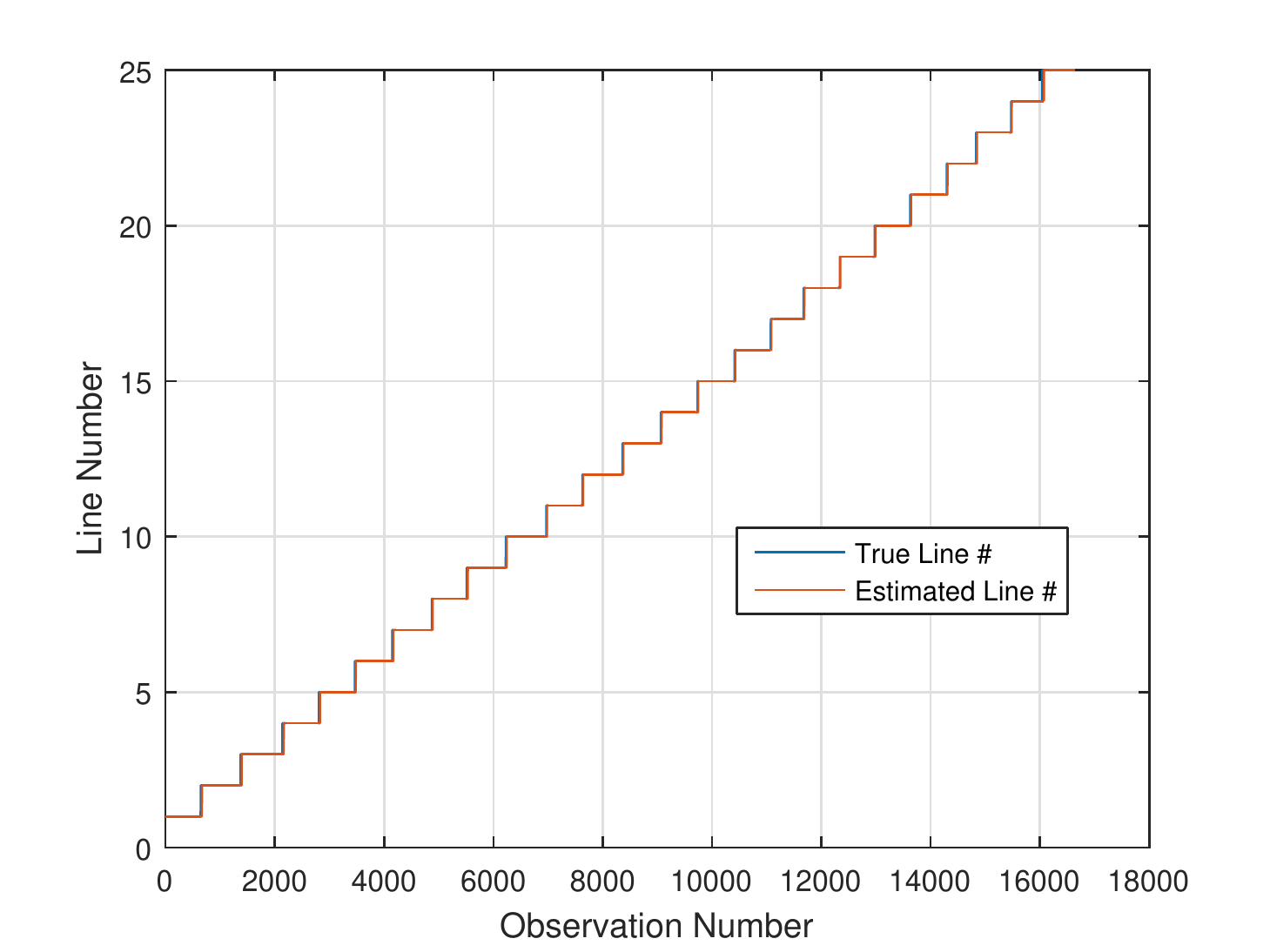} \label{fig:LDSdemo} }\\
\subfloat[][Line detection accuracy for all pages]{\includegraphics[width=.45\textwidth]{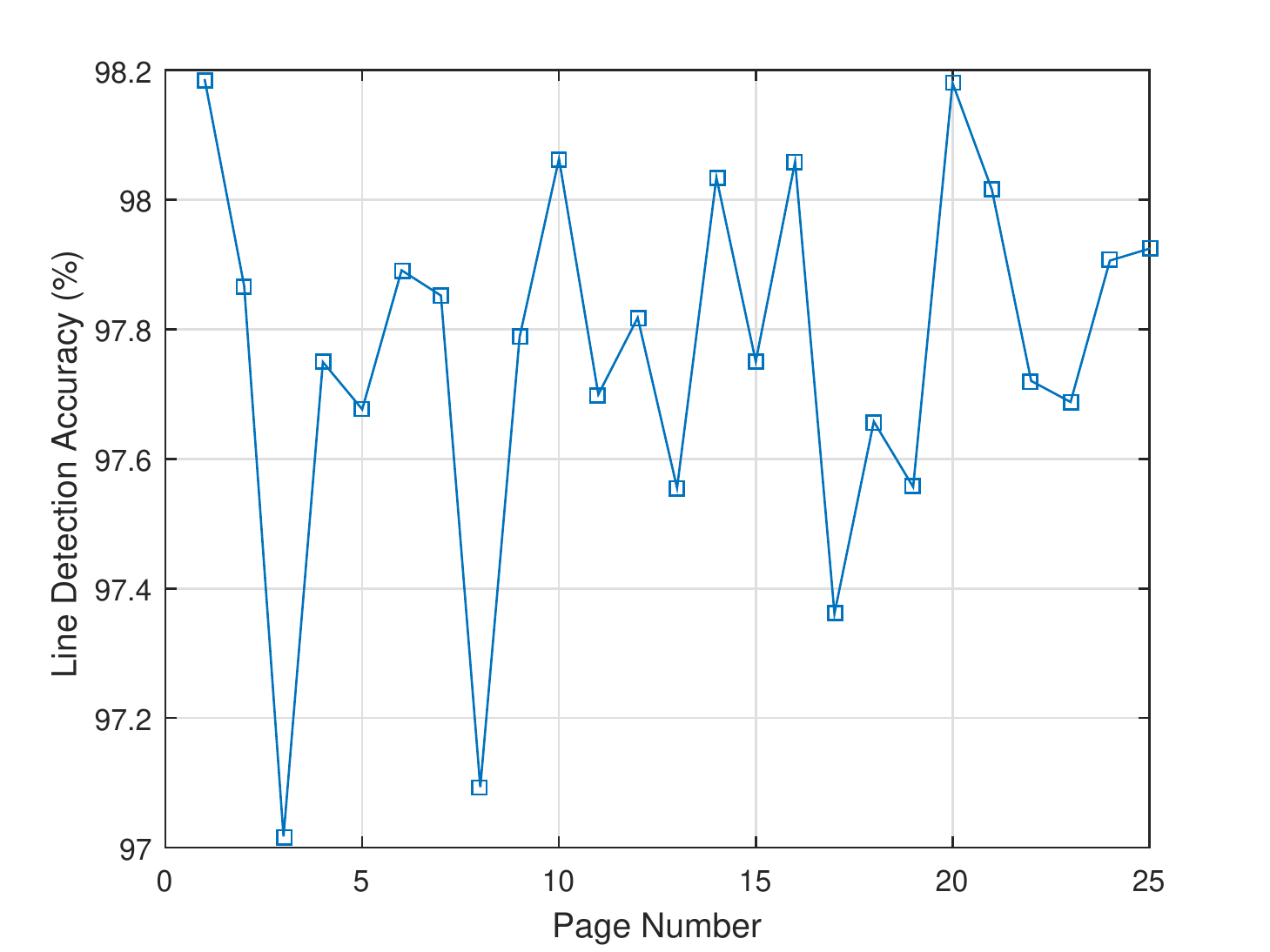} \label{fig:LDSacc}}
\caption{
{\bf Line detection accuracy.}
The line detection accuracy of the proposed method is between 97\% and 98\%. 
}
\label{fig:LDS}
\end{center}
\end{figure}

In terms of line detection, an average {\em line detection accuracy} of 97.83\% was achieved. When compared with the LDS this is an improvement of 8.9\%, using an identical evaluation method as described in \cite{bottos2019approach}. Figure \ref{fig:LDS}\subref{fig:LDSdemo} shows the true line number of each gaze measurement $\bz(k)$ and the estimated line number against $k$ for a single sample page of real-world data, emphasizing the close alignment between the estimated and true line. Figure \ref{fig:LDS}\subref{fig:LDSacc} shows the line detection accuracy per page, for each of the 25 pages of data. The line detection accuracy of the proposed method is consistently between 97\% and 98\%. 

\begin{figure*}[h]
  \begin{center}
  \subfloat[]{\includegraphics[width=.32\textwidth]{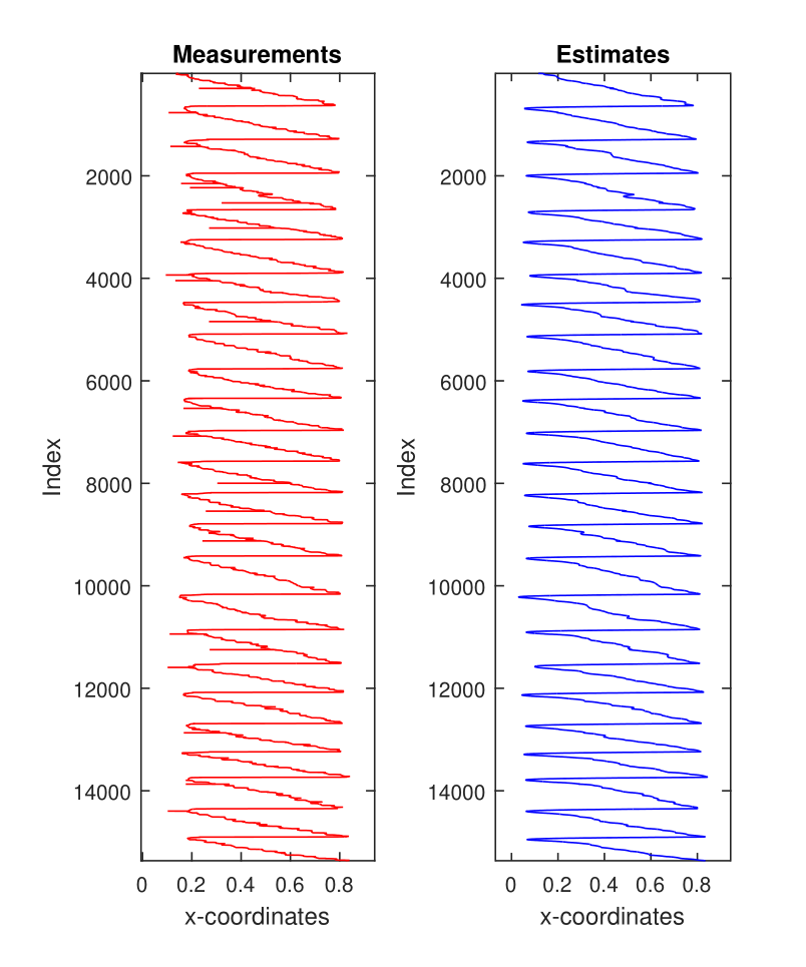} \label{a} } 
  \subfloat[]{\includegraphics[width=.32\textwidth]{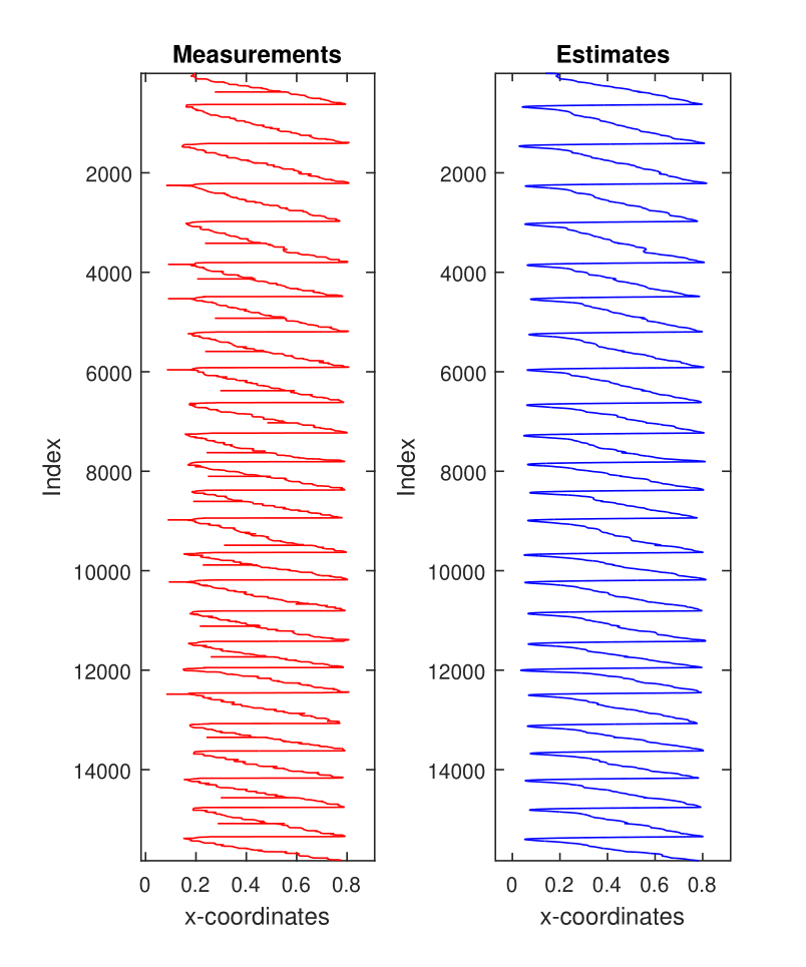} \label{b} }
  \subfloat[]{\includegraphics[width=.32\textwidth]{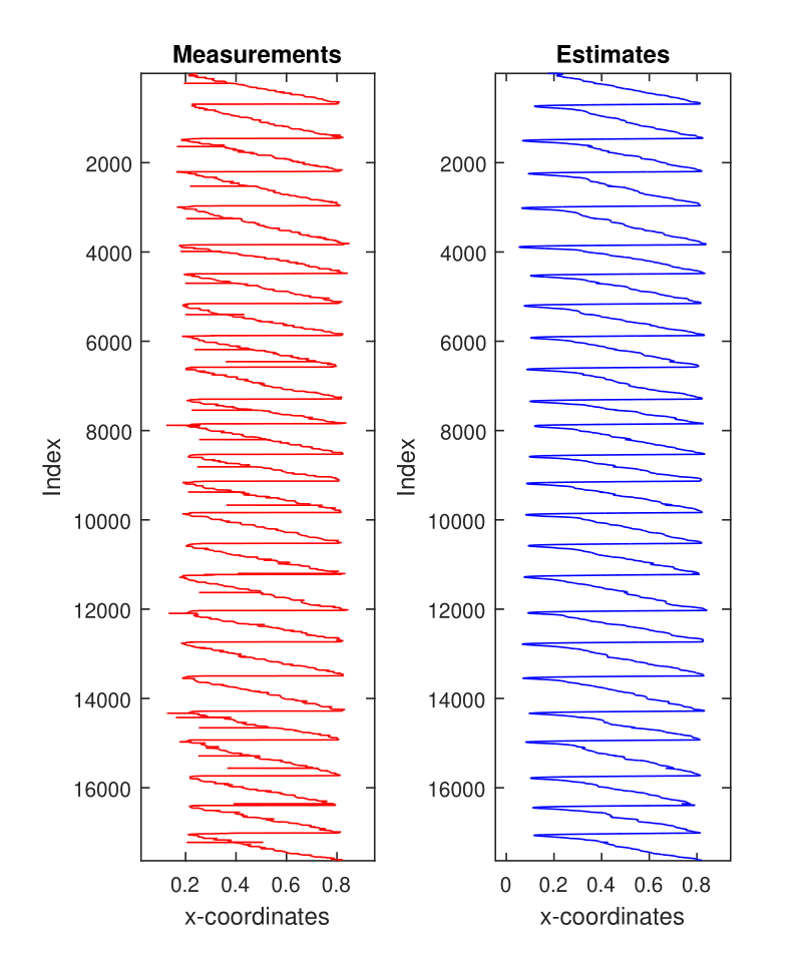} \label{c} }
  \caption{{\bf Comparison between measured and estimated $x$-coordinates.} Three full sample pages of real data are shown, plotting each measured and estimated $x$-coordinate according to their index for comparison. Through examination, one can observe that the Slip-KF estimation procedure is highly effective in terms of noise elimination.
  }
  \label{raw_vs_kf}
  \end{center}
  \end{figure*}  
    
Furthermore, the proposed Slip-KF offers superior noise elimination in the $x$-coordinates when compared with the least squares based approach in \cite{bottosFusion}. Figure \ref{raw_vs_kf} has been included, which illustrates the $x$-coordinates for both the raw (red) and estimated (blue) eye-gaze fixation points plotted side-by-side for comparison, from three full sample pages of real data. Through examination, one can observe that the Slip-KF estimation procedure is highly effective in terms of noise elimination.
  
\section{Conclusions and Discussions}
\label{sec:conclusions}
Let us re-iterate that an effective Slip-Kalman Filter, for the purposes of tracking while reading, must meet the following criteria:
\begin{enumerate}
\item The Slip-Kalman Filter must provide adequate smoothing. For a standard Kalman Filter, too much sensitivity results in the estimated data points mimicking the measured data points too closely, thus providing little effect in terms of noise reduction. The appropriate amount of sensitivity, however, produces the ``slip'' phenomena that was previously discussed.
\item In light of the previous point, the Slip-Kalman Filter must be effective in eliminating the ``slip'' phenomena, allowing the filter to quickly catch up in the event of drastic position and velocity changes.
\item Finally, specific to the work of the authors, the output of the Slip-Kalman Filter must be such that estimated $(x,y)$-coordinates are free of excess noise and are easily interpretable, providing not only a visual picture of an individual's reading progression through a block of text but also an evaluation metric which proves useful in many scenarios - the accurately estimated line being read by an individual at any given time instant, otherwise referred to as ``line-detection accuracy''.
\end{enumerate}
Points 1 and 2 of the above are visually confirmed to be satisfied by inspection of Figures \ref{fig:slipKF} and \ref{raw_vs_kf}. Moreover, the proposed Slip-KF approach demonstrated in this paper, using a commercialy available eye-tracking device to obtain eye-gaze fixation point measurements, achieved a line detection accuracies between 97 and 98\% on a practical dataset where all lines were read once without skipping or repeating. In terms accuracy as well as consistency, detection speed, and effective reduction in noise associated with $(x,\, y)$-coordinates, the Slip-KF is the most effective of all approaches proposed by the authors in recent works, having the added benefit of real-time reading-pattern detection capabilities. 

While the Slip-KF proves to be an effective method of translating raw, noisy eye-gaze fixation points to interpretable data, it has thus far only been tested under the assumption that the reader will not return to previously read text. In other words, each line of text must be read once, and only once, and in sequence. Relaxing this assumption will be the focus of our future work, as an individual's natural reading pattern may invalidate this assumption.

%
%
%
%

\section*{Acknowledgements}
\label{sec:acknowledge}
Dr. Balasingam would like to acknowledge Natural Sciences
and Engineering Research Council of Canada (NSERC) for
financial support under the Discovery Grants (DG) program. 

\bibliography{SlipKF4LDS}

\begin{thebibliography}{10}
\providecommand{\url}[1]{#1}
\csname url@samestyle\endcsname
\providecommand{\newblock}{\relax}
\providecommand{\bibinfo}[2]{#2}
\providecommand{\BIBentrySTDinterwordspacing}{\spaceskip=0pt\relax}
\providecommand{\BIBentryALTinterwordstretchfactor}{4}
\providecommand{\BIBentryALTinterwordspacing}{\spaceskip=\fontdimen2\font plus
\BIBentryALTinterwordstretchfactor\fontdimen3\font minus
  \fontdimen4\font\relax}
\providecommand{\BIBforeignlanguage}[2]{{%
\expandafter\ifx\csname l@#1\endcsname\relax
\typeout{** WARNING: IEEEtranS.bst: No hyphenation pattern has been}%
\typeout{** loaded for the language `#1'. Using the pattern for}%
\typeout{** the default language instead.}%
\else
\language=\csname l@#1\endcsname
\fi
#2}}
\providecommand{\BIBdecl}{\relax}
\BIBdecl

\bibitem{gazept}
``{gazept} gazepoint eye tracker website,'' \url{https://www.gazept.com/},
  accessed: 2018-20-11.

\bibitem{armstrong2012eye}
T.~Armstrong and B.~O. Olatunji, ``Eye tracking of attention in the affective
  disorders: A meta-analytic review and synthesis,'' \emph{Clinical psychology
  review}, vol.~32, no.~8, pp. 704--723, 2012.

\bibitem{bar2004estimation}
Y.~Bar-Shalom, X.~R. Li, and T.~Kirubarajan, \emph{Estimation with applications
  to tracking and navigation: theory algorithms and software}.\hskip 1em plus
  0.5em minus 0.4em\relax John Wiley \& Sons, 2004.

\bibitem{bottos2019approach}
S.~Bottos and B.~Balasingam, ``An approach to track reading progression using
  eye-gaze fixation points,'' \emph{arXiv preprint arXiv:1902.03322}, 2019.

\bibitem{bottosFusion}
------, ``Tracking the progression of reading through eye-gaze measurements,''
  in \emph{22nd International Conference on Information Fusion}.\hskip 1em plus
  0.5em minus 0.4em\relax IEEE, 2019, pp. 1--8.

\bibitem{drieghe2005eye}
D.~Drieghe, K.~Rayner, and A.~Pollatsek, ``Eye movements and word skipping
  during reading revisited.'' \emph{Journal of Experimental Psychology: Human
  Perception and Performance}, vol.~31, no.~5, p. 954, 2005.

\bibitem{hochreiter1997long}
S.~Hochreiter and J.~Schmidhuber, ``Long short-term memory,'' \emph{Neural
  computation}, vol.~9, no.~8, pp. 1735--1780, 1997.

\bibitem{huey1908psychology}
E.~B. Huey, \emph{The psychology and pedagogy of reading}.\hskip 1em plus 0.5em
  minus 0.4em\relax The Macmillan Company, 1908.

\bibitem{hughes1910slip}
J.~W. Hughes, ``Slip-gear.'' Feb.~22 1910, uS Patent 950,290.

\bibitem{klin2002visual}
A.~Klin, W.~Jones, R.~Schultz, F.~Volkmar, and D.~Cohen, ``Visual fixation
  patterns during viewing of naturalistic social situations as predictors of
  social competence in individuals with autism,'' \emph{Archives of general
  psychiatry}, vol.~59, no.~9, pp. 809--816, 2002.

\bibitem{mozaffari2018reading}
S.~S. Mozaffari, F.~Raue, S.~D. Hassanzadeh, S.~Agne, S.~S. Bukhari, and
  A.~Dengel, ``Reading type classification based on generative models and
  bidirectional long short-term memory,'' 2018.

\bibitem{neumann2006looking}
D.~Neumann, M.~L. Spezio, J.~Piven, and R.~Adolphs, ``Looking you in the mouth:
  abnormal gaze in autism resulting from impaired top-down modulation of visual
  attention,'' \emph{Social cognitive and affective neuroscience}, vol.~1,
  no.~3, pp. 194--202, 2006.

\bibitem{paeglis2006maximizing}
R.~Paeglis, K.~Bagucka, N.~Sjakste, and I.~Lacis, ``Maximizing reading: pattern
  analysis to describe points of gaze,'' in \emph{Mathematics of Data/Image
  Pattern Recognition, Compression, and Encryption with Applications IX}, vol.
  6315.\hskip 1em plus 0.5em minus 0.4em\relax International Society for Optics
  and Photonics, 2006, p. 63150R.

\bibitem{rayner1997understanding}
K.~Rayner, ``Understanding eye movements in reading,'' \emph{Scientific Studies
  of Reading}, vol.~1, no.~4, pp. 317--339, 1997.

\bibitem{roper2007louis}
G.~Roper-Hall, ``Louis {\'e}mile javal (1839--1907): The father of
  orthoptics,'' \emph{American Orthoptic Journal}, vol.~57, no.~1, pp.
  131--136, 2007.

\bibitem{sanchez2013attentional}
A.~Sanchez, C.~Vazquez, C.~Marker, J.~LeMoult, and J.~Joormann, ``Attentional
  disengagement predicts stress recovery in depression: An eye-tracking
  study.'' \emph{Journal of Abnormal Psychology}, vol. 122, no.~2, p. 303,
  2013.

\end{thebibliography}

\end{document}